# Influence of the high granularity calorimeter stainless steel absorbers onto the Compact Muon Solenoid inner magnetic field

Vyacheslav Klyukhin[1,2] for the CMS Collaboration

[1*]Skobeltsyn Institute of Nuclear Physics, Lomonosov Moscow State University, RU-119991, Moscow, Russia
[2]CERN, CH-1211, Geneva 23, Switzerland
[*]Corresponding author. E-mail : Vyacheslav.Klyukhin@cern.ch; ORCID: 0000-0002-8577-6531

**Abstract**

The Compact Muon Solenoid (CMS) detector is a general-purpose experimental setup at the Large Hadron Collider (LHC) at CERN to investigate the production of new particles in the proton-proton collisions at a centre of mass energy 13 TeV. The third run of the data taken is started in April 2022 and will continue till the end of 2025. Then, during a long shutdown time, the existing CMS hadron endcap calorimeter will be replaced with a new high granularity calorimeter (HGCal) designed for the higher LHC luminosity. The HGCal contains the stainless-steel absorber plates with a relative permeability limited by a value of 1.05 from estimation of the electromagnetic forces acting on this slightly magnetic material. To exclude the surprises with possible perturbation of the inner magnetic flux density in the region of the charged particle tracking system, an influence of this additional material onto the quality of the magnetic field inside the inner tracker volume is investigated at this limited value of the permeability of stainless steel . The three-dimensional model of the CMS magnet is used for this purpose. The method of the magnetic field double integrals characterizing the charged particle momentum resolution the first time is applied to the CMS detector and the first time is described in the journal publication. The results obtained with this method are presented in detail and discussed.

**Article Highlights**

- The method of the magnetic field double integrals to investigate a quality of the magnetic field in the particle detector is fully described.
- The possible perturbation of the magnetic field inside the Compact Muon Solenoid inner tracker with the replacement of the endcap hadronic calorimeter is investigated.
- The estimations on the degradation of the charged particle momentum resolution are presented.

## 1 Introduction

The Compact Muon Solenoid (CMS) detector [1] at the Large Hardon Collider (LHC) [2] is a multi-purpose experimental setup to register the charged and neutral particles created in the proton-proton collisions at a center of mass energy 13 TeV. The detector includes a wide-aperture superconducting thin solenoid [3] with a diameter of 6 m and a length of 12.5 m operated with a direct current of 18.164 kA to create the central magnetic flux density $B_0$ of 3.81 T. Inside the superconducting coil, the major subdetectors of particles are located around the interaction point of proton beams: a silicon pixel and strip tracking detectors to register the charged particles in a cylinder volume with a diameter of 2.27 m and a length of 5.6 m; a solid crystal electromagnetic calorimeter to register electrons, positrons and gamma rays; a barrel and endcap hadronic calorimeters of total absorption to register the energy of all the particles except of muons and neutrinos. The muons are registered outside the solenoid coil in the muon spectrometer.

After finishing Run 3 of data taken at the LHC in the end of 2025, the existing CMS plastic scintillator-based hadron endcap calorimeter will be replaced by a new High Granularity Calorimeter (HGCal) containing the silicon sensors and stainless-steel absorber plates [4]. The relative permeability of the stainless steel is assuming to be well below 1.05 that is limited by the axial forces attracting the absorber plates to the center of the CMS superconducting solenoid. This study investigates the influence of the HGCal absorber plates with an extreme permeability value of 1.05 onto the CMS solenoid inner magnetic field in the location of the pixel and strip tracking detectors. The study is based on the calculation and comparison of the magnetic field double integrals characterizing the resolution in the charged particle momentum measurements.

The method of the magnetic field double integrals has been proposed in 1993 [5] and was used in the investigation of the magnetic field quality in the different options [6–8] of the magnetic system for the FCC-hh detector [9] at the proposing Future Circular Collider [10].

The structure of this article is as follows: in Section 2 the method of the magnetic field double integrals applied to the CMS inner magnetic field is described in detail; in Section 3 the model of the HGCal stainless steel absorber plates is presented and the expressions for the axial force calculations are displayed; in Section 4 the results of the study are drawn and discussed, and finally, Section 5 contains a conclusion.

## 2 Magnetic field double integrals and particle momentum resolution

In a center of mass reference system of the colliding particle beams, consider a trajectory of a charged particle emitted in the radial *RZ*-plane, at angle $\theta$ to the beam *Z*-axis, from the nominal beam crossing point. For a small step *dl* along the direction of the particle motion in an ideal (homogeneous) solenoid the change to the turning angle of the track $d\alpha$ lies in the transverse plane and is given by

$$d\alpha = \frac{0.3}{p_T} B \, dl \, sin\theta, \qquad (1)$$

where *l* is in meters, $p_T$ is the particle transverse momentum of the constant value in GeV/*c*, and **B** is the constant vector of the magnetic flux density in Tesla. In general, for inhomogeneous field, where the vector **B** changes its value and direction, the track is turning according to, and in the direction of, the vector product *dl*×**B** [5].

For energetic particles, the magnetic deflection is small compared to the track length, thus the distance along the trajectory can be approximated by $l = r / sin\theta$, where *r* is the transverse radius (the orthogonal distance from the beam axis to *l* in the radial *RZ*-plane), and small angle approximations are valid. At a track length *l* in the *RZ*-plane, the relative angle $\alpha$ of the track with respect to its initial direction in transverse projection is given by

$$\alpha(l) = \frac{0.3}{p_T} \int_0^l B \, sin\theta_{(dl,B)} \, dl. \qquad (2)$$

Here the polar angle $\theta_{(dl, B)}$ represents the longitudinal component of the angle between the track projection to the *RZ*-plane and the field vector, i. e. both the track length and the magnetic flux density vector are considered to lay in the *RZ*-plane.

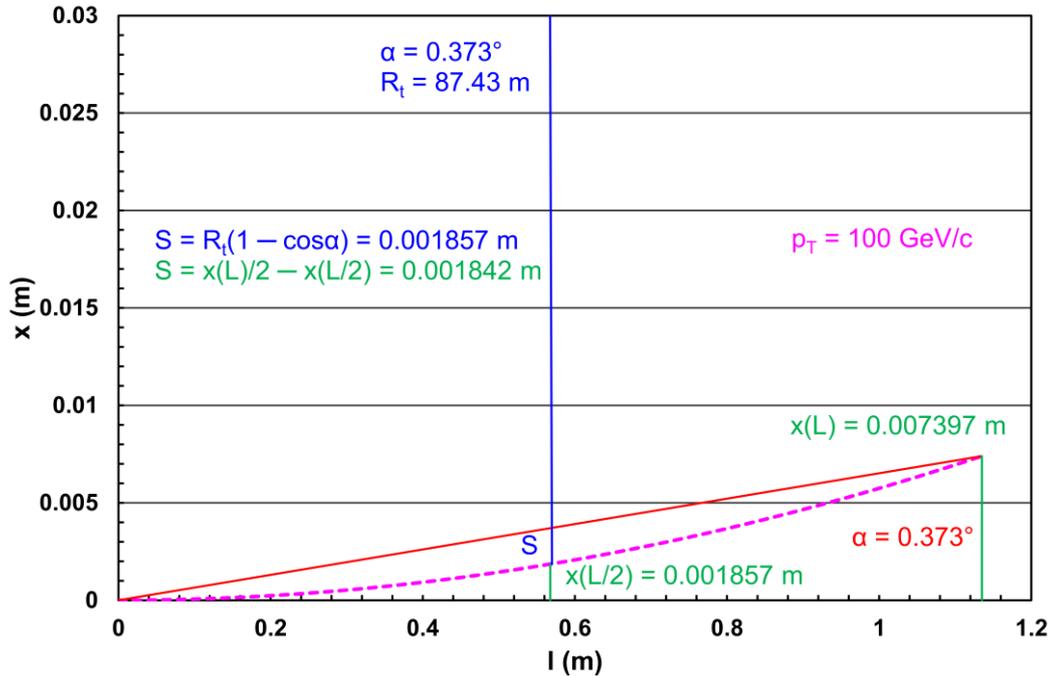

**Fig. 1:** The trajectory of a charged particle (magenta dashed line) with a transverse momentum $p_T$ of 100 GeV/*c* emitted in a plane transverse to the axis of the interacting particle beams. Here: $R_t$ (blue line) is a radius of the particle trajectory; *S* (an end part of blue line) is a distance from the center of the arc to the center of its base (sagitta); $\alpha$ (red line) is a particle final turning angle; transverse deflection *x* (green lines) is determined with Eq. (3); *L* is the total track length in the *RZ*-plane. The magnetic flux density vectors are orthogonal to the plane of the plot and correspond to the distribution of the magnetic flux in the existing CMS configuration. The sagitta *S* is calculated in two ways: with the exact formula (in blue) and with the transverse deflections *x* (in green). The difference between two calculated values is within 0.8 %. The scale of the vertical axis distorts the trajectory radius turning angle, half of which is also equal to $\alpha$.

The total transverse deflection $x$ shown in Fig. 1 is obtained by integration Eq. (2) over $dr = dl\, sin\theta$:

$$x(l) = \frac{0.3}{p_T} \int_0^{l\, sin\theta} \int_0^{r/sin\theta} B\, sin\theta_{(dl,B)}\, dl\, dr. \qquad (3)$$

For the ideal solenoid, $x(l)$ is simply proportional to $Bl^2$.

In the silicon tracker, the momentum resolution at the large transverse momenta $p_T$ [GeV/$c$] is dominated by the detector resolution, thus the relative transverse momentum precision $\delta$ in the ideal solenoid can be approximated by the Gluckstern formulae [11, 9]:

$$\delta = \frac{dp_T}{p_T} \approx \frac{\sigma p_T}{0.3 BL^2}\sqrt{\frac{720}{N+4}}, \qquad (4)$$

expressed for a uniform solenoid field $B$ [T], $N$ equidistant detector planes, a resolution $\sigma$ [m] per layer, and a track length of $L$ [m] in the tracker volume. In the inhomogeneous magnetic field, the term $BL^2$ should be replaced by the magnetic field double integral

$$I_2 = \int_0^{l\, sin\theta} \int_0^{r/sin\theta} B\, sin\theta_{(dl,B)}\, dl\, dr. \qquad (5)$$

To provide the precise charged particle momentum measurements, the track sagitta can be approximated as $x(L)/2 - x(L/2)$ as shown in Fig. 1, where $L$ is the total track length in the $RZ$-plane of the tracking volume. For the ideal solenoid, the sagitta equals $x(L/2)$ since $x(L) = 4x(L/2)$ from Eq. (3).

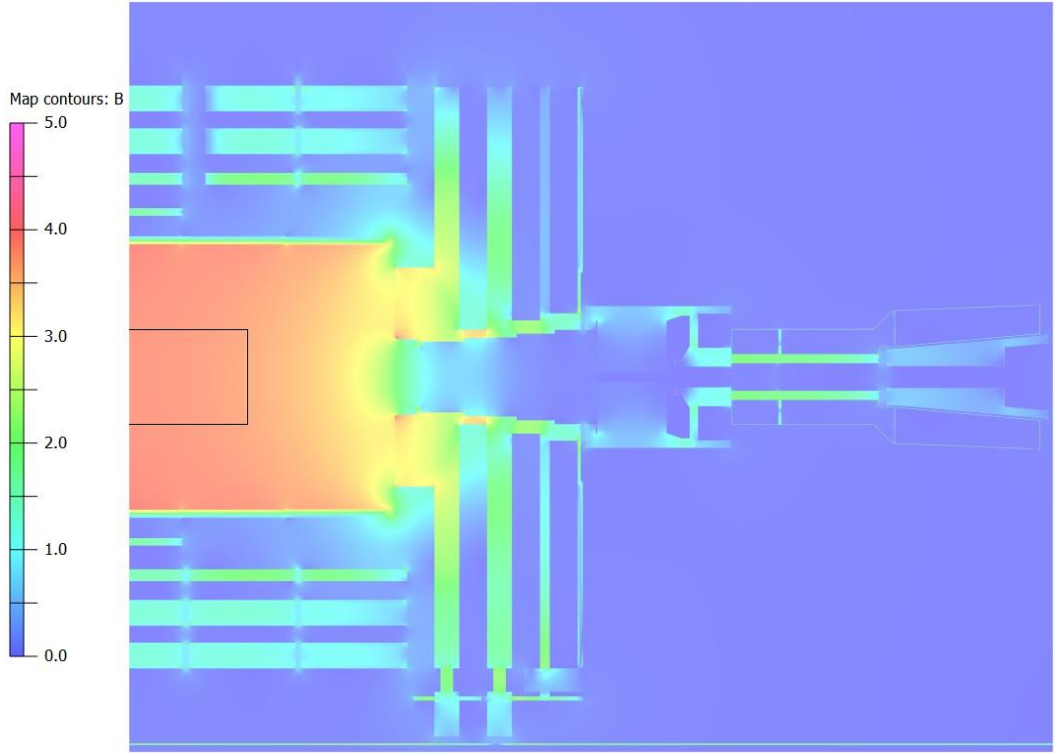

**Fig. 2:** A half of the present CMS magnetic field map prepared in the longitudinal section with the latest CMS magnet three-dimensional (3D) model [12]. The color scale of the calculated total magnetic flux density $B$ is from zero to 5 T with an increment of 0.5 T. A black rectangular in the middle of the solenoid coil represents a half of the active tracker volume with half an axial length of 2.8 m and a diameter of 2.27 m.

From Eqs. (3) and (4) the relative momentum precision for a real versus ideal solenoid is given by a ratio $R = \delta_h / \delta_i$, or $R = x_i(L) / x_h(L)$, or $R = I_{2i} / I_{2h}$, where indexes $h$ and $i$ stand for the homogeneous and inhomogeneous magnetic field, accordingly. The degradation of the charged particle relative transverse momentum precision is proportional to 1 − R. The track length L depends on the pseudorapidity η determined by $\eta = -ln[tan(\theta/2)]$, where θ is a polar angle of particle in the reference frame. The value $\eta_c$ = 1.63489 corresponds to the corner of the CMS inner tracker volume in the RZ-plane. For the pseudorapidity values smaller than $\eta_c$, the length L is equal to $R_{max} / sin\theta$, for the larger values the length L is equal to $Z_{max} / cos\theta$, where $R_{max}$ and $Z_{max}$ are the radius and half a length of the active inner tracker volume shown in Fig 2 by the black rectangular.

## 3 Description of the HGCal absorber plate model

To absorb the hadronic particles along the entire length of the HGCal, 22 stainless steel plates with thicknesses from 45 to 60.7 mm interleaved with the air gaps of 21.55 mm will be located near both ends of the superconducting coil at the distances from 3.6098 to 5.2239 m from the coil middle plane. In these distances the shift by 12 mm to the coil center caused by the magnet yoke deformation under the magnetic forces is considered.

As shown in Figs. 3 and 4, the plates form two cones with a small slope to the axis and a large slope to the axis, and two cylinders at each side of the coil inner volume. The smallest outer diameter of the small slope cone is 3.2934 m, the outer diameter of the large cylinder is 5.2492 m, and the volume of absorber plates in each HGCal endcap is about 21 m$^3$. The cylinder of the active inner tracker volume of 2.27 m in diameter and 5.6 m in length is located between the HGCal endcaps.

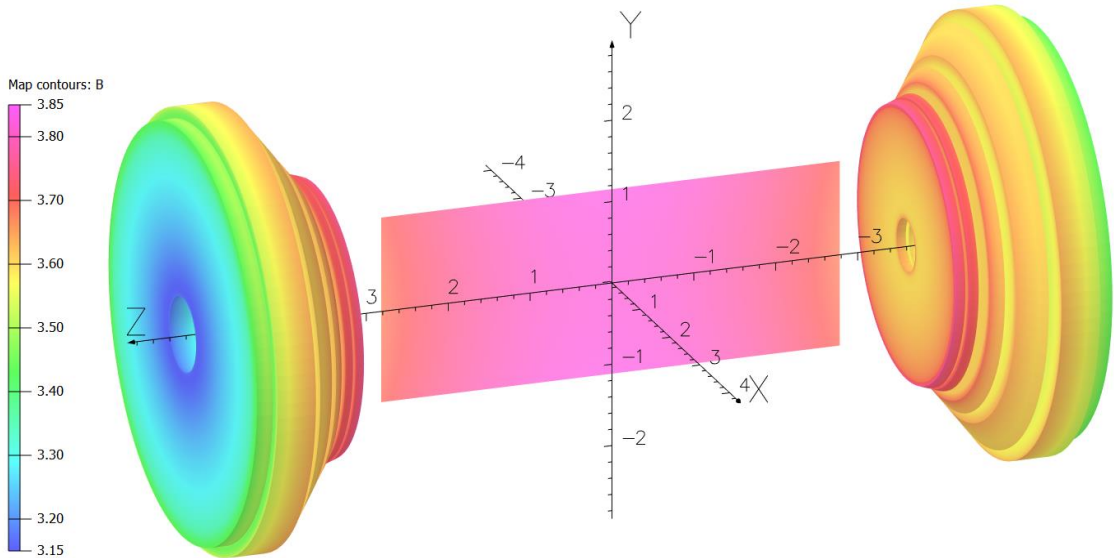

**Fig. 3:** 3D model of the HGCal absorber plates with a relative permeability of 1.05. The calculated magnetic flux density B on the absorber surfaces varies from 3.75 to 3.15 T (−15.8 %). A rectangular plot in the middle represents the magnetic flux density across the tracker vertical RZ-plane with a variation from 3.83 to 3.68 T (−3.97 %). The color scale is in Tesla with a unit of 0.1 T.

Both Figs. 3 and 4 are prepared with the updated latest CMS magnet 3D model [12]. To avoid the substantial change of the model with description of all the 22 absorber plates in each endcap, a simplified geometry of the absorbers is introduced in the model. As shown in Fig. 4, each endcap is presented in a form of five sections with four air gaps between them. One air gap is in the small slope cone, three air gaps are in the large slope cone, the volume of stainless steel in each modelled endcap is 21.4 m$^3$. The model of the CMS magnet is calculated with the operational direct current of 18.164 kA and with the value of the stainless steel relative permeability 1.05. The coordinate axis shown in Figs. 3 and 4 represent the CMS coordinate system where the origin is in the center of the superconducting solenoid, the X axis lies in the LHC plane and is directed to the center of the LHC machine, the Y axis is directed upward and is perpendicular to the LHC plane, the Z axis makes up the right triplet with the X and Y axes and is directed along the vector of magnetic flux density created on the axis of the superconducting coil.

The value of the permeability used in the calculations corresponds to the limitation of the axial magnetic forces onto the absorbers within certain limits. In the model, the axial forces at a relative permeability $\mu_{rel}$ equal to 1.05 are calculated to be −1.087 MN per the modelled endcap with positive Z-coordinates, and +1.082 MN per the modelled endcap with negative Z-coordinates. The difference arises due to the absence of one turn in the superconducting coil module located on the negative Z coordinates [12].

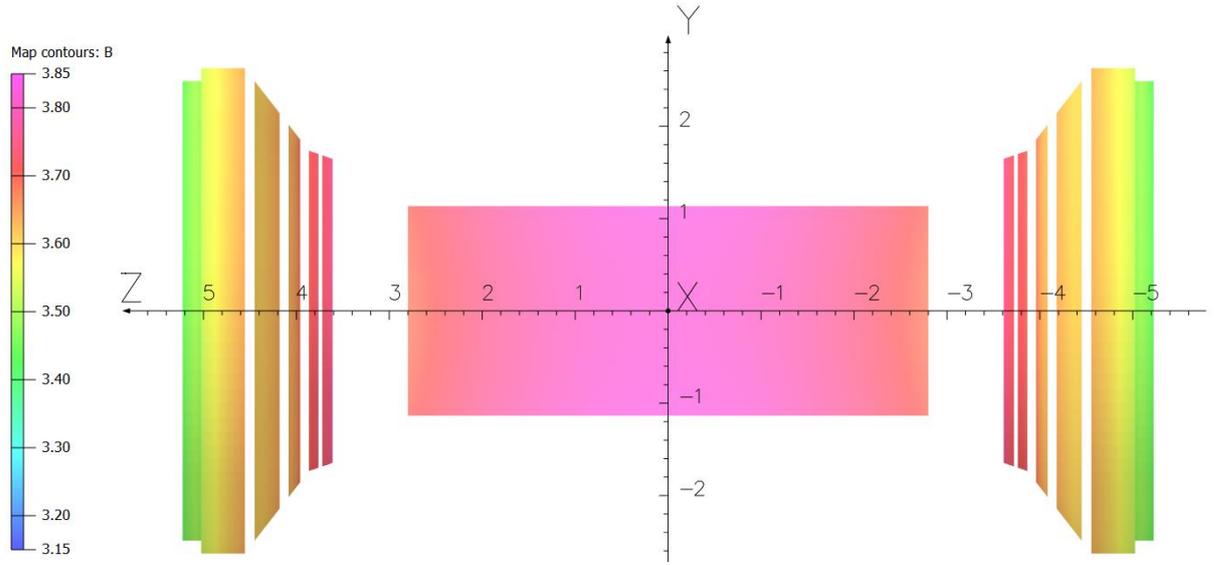

**Fig. 4:** Projection of the inner tracker volume of 5.6 m long and 2.27 m in diameter, and the HGCal absorber plates onto the longitudinal section of the CMS coil inner volume. The central magnetic flux density $B_0$ is 0.228 % greater than that value in the existing CMS configuration. The color scale is in Tesla with a unit of 0.1 T.

Based on the observed linear dependence of the axial force on permeability values below 1.1, and considering the zero value of the axial force at permeability of 1.0, the following expressions for axial forces are valid:

$$F_{z+} [MN] = -21.74 \cdot (\mu_{rel} - 1), \qquad (6)$$
$$F_{z-} [MN] = +21.64 \cdot (\mu_{rel} - 1), \qquad (7)$$

where $\mu_{rel}$ is the relative permeability in the range from 1.0 to 1.05, $F_{z+}$ is the axial force in MN on the absorber plates at positive Z-coordinates, and $F_{z-}$ is the axial force in MN on the absorber plates at negative Z-coordinates. Both forces are directed towards the center of the solenoid coil.

The goal of reducing axial forces by using the lowest possible values of relative permeability in stainless steel is achievable [13]. To consider the spread of the permeability values across the plates, all the 22 plates in each endcap should be described in the next version of the CMS magnet model.

## 4 Results and discussion

The magnetic field double integrals $I_2$ from Eq. (5) are calculated in a quarter of the inner tracking volume vertical *RZ*-plane in the pseudorapidity range from 0 to 3. The further comparisons are performed for the variables as follows: a ratio $R_{ih} = I_{2i} / I_{2h}$, a ratio $R_{ah} = I_{2a} / I_{2h}$, and a ratio $R_{ai} = I_{2a} / I_{2i}$, where indexes *h*, *i*, and *a* stand for the homogeneous magnetic field, the present inhomogeneous CMS field, and the magnetic field affected by the stainless steel absorbers, accordingly. The ideal (homogeneous) solenoid is assumed to be with the constant magnetic flux density $B$ of 3.809442 T, corresponded to the central field in the existing CMS configuration.

In Fig. 5 the dependence of the total magnetic flux density on the track length at different pseudorapidity values is displayed in the CMS inner tracking volume affected by the HGCal stainless steel absorbers. The central magnetic flux density $B_0$ is 0.228 % greater than in the existing CMS configuration. For small values of the pseudorapidity the total magnetic flux density $B$ rises with the track length and for $\eta > 0.65$ falls. This behavior has not changed in comparison with the present CMS magnetic field, and the values of the magnetic flux density are almost the same.

In Fig. 6 the angle between the total magnetic flux density vector and the track direction along the track length is displayed at different pseudorapidity values in the CMS inner tracking volume affected by the HGCal stainless steel absorbers. In both CMS configurations (with and without the stainless steel absorbers) the variation of this angle along the track length is extremely small.

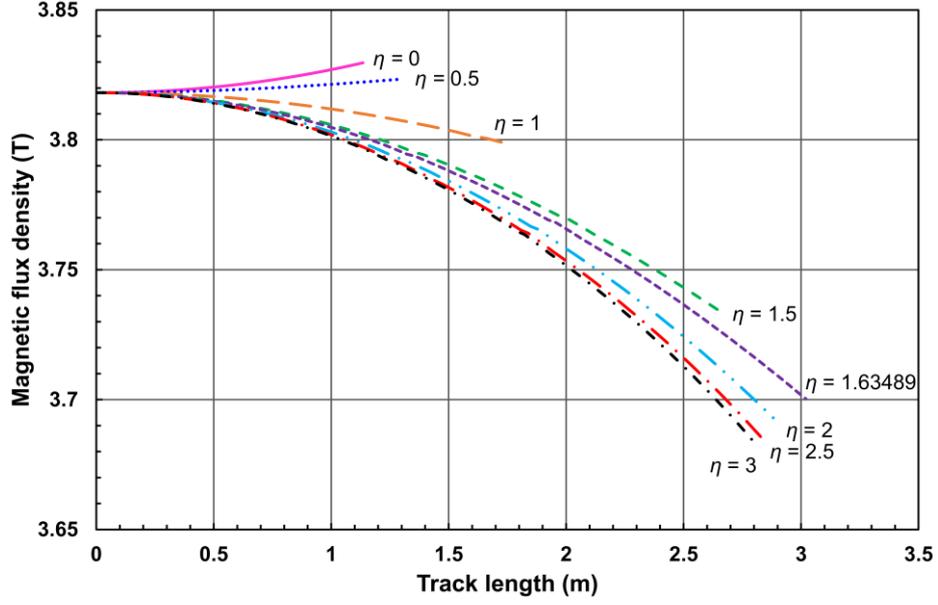

**Fig. 5:** Magnetic flux density along the track length at different values of the pseudorapidity in the vertical plane of the CMS inner tracking volume affected by the HGCal absorber plates.

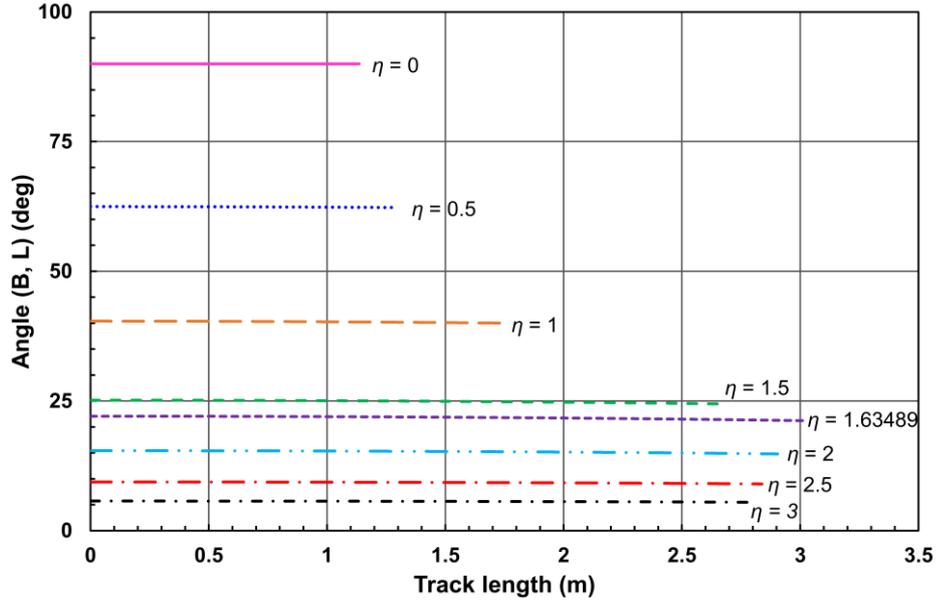

**Fig. 6:** The angle between the magnetic flux density vector and the track direction along the track length at different values of the pseudorapidity in the vertical plane of the CMS inner tracking volume affected by the HGCal absorber plates.

In Fig. 7 the dependence of all three double integrals $I_{2h}$, $I_{2i}$, and $I_{2a}$ from Eq. (5) on the pseudorapidity is presented. For $\eta < \eta_c$, the integral $I_{2h}$ drops from 2.4645 to 2.4578 T·m², the integral $I_{2i}$ drops from 2.4658 to 2.4271 T·m², and the integral $I_{2a}$ drops from 2.4713 to 2.4361 T·m². In the pseudorapidity interval $\eta > \eta_c$, all three integrals drop rapidly by restriction of the track length with $Z_{max}$ and by decreasing the angle between the total magnetic flux density vector and the track direction at the large values of pseudorapidity. In all the pseudorapidity range from 0 to 3 the ratios $R$ stay in the limits as follows: $1.0006 > R_{ih} > 0.9869$, $1.0028 > R_{ah} > 0.9906$, $1.0022 > R_{ai} > 1.0037$. These results show that both integrals $I_{2i}$ and $I_{2a}$ degrades with pseudorapidity comparing with the ideal magnetic field double integral $I_{2h}$, but the integral $I_{2a}$ is always larger than integral $I_{2i}$ by 0.22 − 0.37 %. That means that the insertion of the HGCal absorbers does not impair the magnetic field and the momentum resolution in the inner tracking volume.

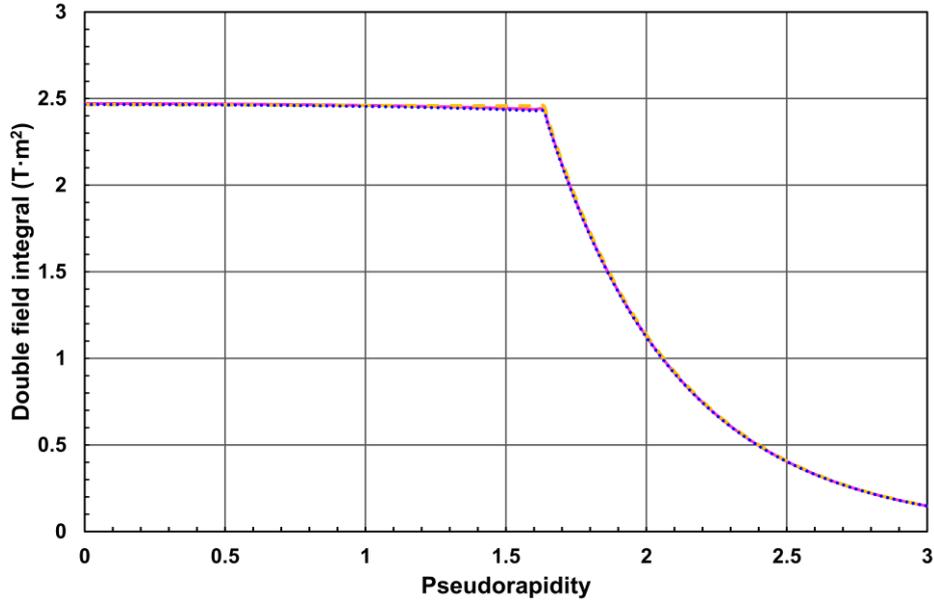

**Fig. 7:** The magnetic field double integrals $I_{2h}$ (dashed line), $I_{2i}$ (dotted line), and $I_{2a}$ (solid line) vs. pseudorapidity in the vertical plane of the CMS inner tracking volume.

In Fig. 8 the magnetic field double integral degradation $1-R$ with respect to homogeneous and inhomogeneous field integrals is shown for three cases: $1-R_{ih}$, $1-R_{ah}$, and $1-R_{ai}$. For the pseudorapidity interval $\eta > \eta_c$, the degradation of the integrals varies slightly as follows: $0.0125 < 1-R_{ih} < 0.0131$, $0.00883 < 1-R_{ah} < 0.00937$, and $-0.00368 > 1-R_{ai} > -0.00374$.

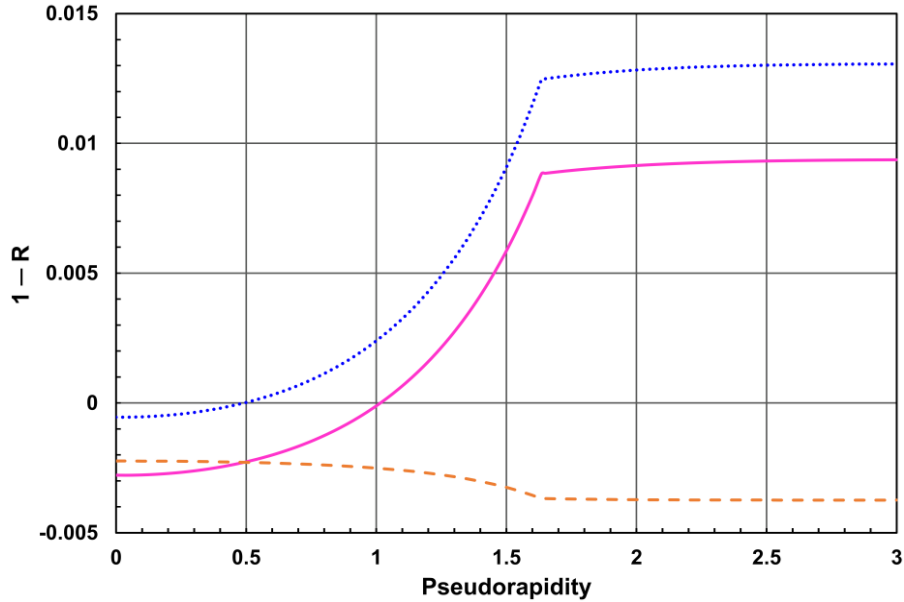

**Fig. 8:** The magnetic field double integral degradation $1-R_{ih}$ (dotted line), $1-R_{ah}$ (solid line), and $1-R_{ai}$ (dashed line) vs. pseudorapidity in the vertical plane of the CMS inner tracking volume.

These results clearly show that the CMS magnetic field inside the inner tracking volume, affected or not by the HGCal absorber plates, is rather homogeneous and thus, provides the high momentum resolution in measuring the trajectories of the charged particles in both cases.

The application of the magnetic field double integral method [5] to the investigation of the CMS inner tracker magnetic field quality allows to conclude without any time consuming simulations that the insertion of the HGCal stainless steel absorber plates with relative permeability of 1.05 does not degrade the momentum resolution of charged particles. The geometric parameters of the CMS solenoid with a coil diameter to coil length ratio of about 0.5 provide a high-quality magnetic field in the active tracking volume with a diameter 2.8 times less than the coil diameter and a length 2.2 times less than the length of the coil. The high quality of this field is not compromised by inserting about 21 m$^3$ of the slightly magnetic stainless steel material at 0.8 m from each end of the tracker volume.

## 5 Conclusion

The method of the magnetic field double integrals singles out the effect of the quality of the magnetic field on the momentum resolution of the charged particles. This method is used to study the perturbation of the CMS inner magnetic field by the high granularity calorimeter stainless steel absorbers with relative permeability of 1.05. The central magnetic flux density is increased by 0.228 % compared to the existing CMS configuration. The maximum magnetic field double integral degradation inside the inner tracker volume improves from 1.3 % to 0.9 %. The small effect of the high granularity calorimeter insertion on the homogeneity of the magnetic flux density in the tracking volume allows precise measurements of the momenta of the charged particles to be maintained with the present description of the CMS magnetic field map.

## Declarations

*Author Contributions:* The author contributed to the study conception and design. Material preparation, data collection and analysis were performed by Vyacheslav Klyukhin. The first draft of the manuscript was written by Vyacheslav Klyukhin who read and approved the final manuscript.
*Funding:* The author declare that no funds, grants, or other support were received during the preparation of this manuscript.
*Data availability:* The author declare that manuscript has no associated data.
*Conflict of Interest:* Author declares that he has no conflict of interest.
*Ethical approval:* This article does not contain any studies with human participants or animals performed by the author.## References

[1] CMS Collaboration (2008) The CMS experiment at the CERN LHC. JINST **3**, S08004. https://doi.org/10.1088/1748-0221/3/08/S08004.
[2] LHC Machine (2008) Editors: Evans, L., Bryant, P. JINST **3**, S08001. https://doi.org/10.1088/1748-0221/3/08/S08001.
[3] Hervé, A. (2010) Constructing a 4-Tesla large thin solenoid at the limit of what can be safely operated. Mod. Phys. Lett. **A25**, 1647–1666. https://doi.org/10.1142/S0217732310033694
[4] CMS Collaboration (2018) The Phase-2 Upgrade of the CMS endcap calorimeter. Technical Design Report. CERN-LHCC-2017-023, CMS-TDR-019, pp. 11–20. CERN, Geneva (2018). ISBN: 978-92-9083-459-5.
[5] Klyukhin, V.I., Poppleton, A., Schmitz, J. (1993) Field integrals for the ATLAS tracking volume. Preprint at https://arxiv.org/abs/1808.00955.
[6] Bielert, E.R., Berriaud, C., Curé, B., Dudarev A., et al. (2016) Design of a 56-GJ Twin Solenoid and Dipoles Detector Magnet System for the Future Circular Collider, IEEE Trans. Appl. Supercond. **26**, 4003506. https://doi.org/10.1109/TASC.2016.2528988.
[7] Mentink, M., Dudarev, A., Pais Da Silva, H.F., Rolando, G., et al. (2016) Iron-free detector magnet options for the future circular collider, Phys. Rev. Accel. Beams **19**, 111001. https://doi.org/10.1103/PhysRevAccelBeams.19.111001.
[8] Bielert, E.R., Berriaud, C., Curé, B., Dudarev A., et al. (2018) Design of the Optional Forward Superconducting Dipole Magnet for the FCC-hh Detector, IEEE Trans. Appl. Supercond. **28**, 4500204. https://doi.org/10.1109/TASC.2018.2813535.
[9] Abada, A., Abbrescia, M., AbdusSalam, S.S., Abdyukhanov I., et.al. (2019) FCC-hh: The Hadron Collider. Future Circular Collider Conceptual Design Report Volume 3, Eur. Phys. J. Special Topics **228**, 755–1107, p. 964. https://doi.org/10.1140/epjst/e2019-900087-0.
[10] Abada, A., Abbrescia, M., AbdusSalam, S.S., Abdyukhanov I., et.al. (2019) FCC Physics Opportunities. Future Circular Collider Conceptual Design Report Volume 1, Eur. Phys. J. C **79**, 474. https://doi.org/10.1140/epjc/s10052-019-6904-3.
[11] Gluckstern, R.L. (1963) Uncertainties in track momentum and direction, due to multiple scattering and measurement errors, Nuclear Instruments and Methods **24**, 381–389. https://doi.org/10.1016/0029-554X(63)90347-1.
[12] Klyukhin, V. (2021) Design and Description of the CMS Magnetic System Model. Symmetry **13**, 1052. https://doi.org/10.3390/sym13061052
[13] Wilson, N., Bunch, P. (1991) Magnetic permeability of stainless steel for use in accelerator beam transport systems. In Proceedings of the 1991 IEEE Particle Accelerator Conference, San Francisco, CA, USA, 6–9 May 1991; pp. 2322–2324. https://doi.org/10.1109/PAC.1991.164953.8